\documentclass[12pt]{iopart}
\usepackage{graphicx}
\begin{document}

\title{Quenched bond randomness in marginal and non-marginal Ising spin models in 2D}

\author{N G Fytas, A Malakis and I A Hadjiagapiou}

\address{Department of Physics, Section of Solid State
Physics, University of Athens, Panepistimiopolis, GR 15784
Zografos, Athens,
Greece}\eads{\mailto{nfytas@phys.uoa.gr},\mailto{amalakis@phys.uoa.gr},\mailto{ihatziag@phys.uoa.gr}}

\begin{abstract}
We investigate and contrast, via entropic sampling based on the
Wang-Landau algorithm, the effects of quenched bond randomness on
the critical behavior of two Ising spin models in 2D. The random
bond version of the superantiferromagnetic (SAF) square model with
nearest- and next-nearest-neighbor competing interactions and the
corresponding version of the simple Ising model are studied and
their general universality aspects are inspected by a detailed
finite-size scaling (FSS) analysis. We find that, the random bond
SAF model obeys weak universality, hyperscaling, and exhibits a
strong saturating behavior of the specific heat due to the
competing nature of interactions. On the other hand, for the
random Ising model we encounter some difficulties for a definite
discrimination between the two well-known scenarios of the
logarithmic corrections versus the weak universality. Yet, a
careful FSS analysis of our data favors the field-theoretically
predicted logarithmic corrections.
\end{abstract}

\vspace{2pc} \noindent{\it Keywords}: Classical Monte Carlo
simulations, Classical phase transitions (Theory), Finite-size
scaling, Disordered systems (Theory) \maketitle

\section{Introduction}
\label{sec:1}

The understanding of the role played by impurities on the nature
of phase transitions is of great importance, both from
experimental and theoretical perspectives. First-order phase
transitions are known to be dramatically softened under the
presence of quenched
randomness~\cite{imry-79,aizenman-89,hui-89,chen-95,chatelain-01,fernandez-08},
while continuous transitions may have their exponents altered
under random fields or random
bonds~\cite{hui-89,harris-74,chayes-86}. There are some very
useful phenomenological arguments and some, perturbative in
nature, theoretical results, pertaining to the occurrence and
nature of phase transitions under the presence of quenched
randomness~\cite{hui-89,dotsenko-95,cardy-96,jacobsen-98,olson-00,chatelain-00}.
The most celebrated criterion is that suggested by
Harris~\cite{harris-74}. This criterion relates directly the
persistence, under random bonds, of the non random behavior to the
specific heat exponent $\alpha_{p}$ of the pure system. According
to this criterion, if $a_{p}$ is positive, then the disorder will
be relevant, i.e., under the effect of the disorder, the system
will reach a new critical behavior. Otherwise, if $a_{p}$ is
negative, disorder is irrelevant and the critical behavior will
not change. The value $\alpha_{p}=0$ is an inconclusive, marginal
case. The 2D Ising model falls into this category, it is the most
studied case, but is still controversial~\cite{MK-99}. In general
and despite the intense efforts of the last years on several
different models, our current understanding of the quenched
randomness effects is rather limited and the situation appears
still unclear for both cases of first- and second-order phase
transitions. The present paper aims to contribute to our knowledge
of the effects of quenched bond disorder on second-order phase
transitions. In this quite active field of research, the resort to
large scale Monte Carlo simulations is often necessary and useful,
and following this recipe, we applied some recently developed by
our group numerical schemes - see reference~\cite{fytas-08a} and
references therein - on two types of random bond Ising models.

In the first part of the paper we present an extension of our
numerical investigation of a random bond spin model in 2D with
competing interactions, known as the random bond square SAF
model~\cite{fytas-08b}. Details of the pure and random version of
this model with competing interactions and the motivation of such
a study will be given below in the corresponding Section, but can
also be found in our recently published Letter~\cite{fytas-08b}
and in other related papers in the literature~\cite{butera-08}.
Furthermore, in the second part of our work, by a parallel study -
using the same numerical techniques - we attempt to shed new light
into the well-known random bond version of the 2D (simple) Ising
model. Our investigation will be related to the extensive relevant
literature concerning this
case~\cite{DD-81,Shalaev-84,Cardy-86,Shankar-87,Ludwig-87,LC-87,Mayer-89,wang-90,Ludwig-90,
Ziegler-90,WSDA-90,Heuer-92,Shalaev-94,KP-94,Kuhn-94,QS-94,TS-94,
MS-95,AQS-96,JS-96,CHMP-97,BFMMPR-97,AQS-97,SSLI-97,
RAJ-98,SSV-98,AQS-99,LSZ-01,Nobre-01,SV-01,TO-01,MC-02,
COPS-04,Queiroz-06,LQ-06,PHP-06,kenna-06,MP-07,hasenbusch-08,hadjiagapiou-08,kenna-08}.
In particular, our discussion will focus on the main point of the
last two decades, concerning the two well-known conflicting
scenarios, namely the logarithmic
corrections~\cite{DD-81,Shalaev-84,Shankar-87,Ludwig-87} versus
the weak universality
scenario~\cite{KP-94,Kuhn-94,suzuki-74,gunton-75}.

The rest of the paper is organized as follows: In the next Section
we outline an extensive entropic sampling program. This program is
based on (i) the Wang-Landau (WL) method~\cite{WL-01}, (ii) the
dominant energy restriction scheme~\cite{malakis-04}, and (iii) a
second stage improvement that combines the WL method~\cite{WL-01}
and some new ideas~\cite{fytas-08a,fytas-08b,BP-07}, suitable for
the study also of disordered systems. Our FSS analysis of the
numerical data and the corresponding discussion of the random bond
versions of the square SAF model and the 2D Ising model are
presented in Sections~\ref{sec:3} and \ref{sec:4}, respectively.
Finally, our conclusions are summarized in Section~\ref{sec:5}.

\section{Simulation schemes and numerical details}
\label{sec:2}

Importance sampling methods have been for many years the main
tools in condensed matter physics and critical
phenomena~\cite{metropolis-53,bortz-75,binder-97,newman-99,landau-00}.
However, for complex systems, effective potentials may have a
rugged landscape, that becomes more pronounced with increasing
system size. In such cases, these traditional methods become
inefficient, since they cannot overcome such barriers in the state
space. A large number of generalized ensemble methods have been
proposed to overcome such
problems~\cite{WL-01,newman-99,landau-00,lee-93,lee-06,oliveira-96,wang-99,berg-92,smith-95,torrie-97,
swendsen-86,geyer-91,marinari-92,lyubartsev-92,hukushima-96,marinari-98,trebst-04}.
One important class of these methods emphasizes the idea of
directly sampling the energy density of states (DOS) and may be
called entropic sampling methods~\cite{newman-99}. In entropic
sampling, instead of sampling microstates with probability
proportional to $e^{-\beta E}$, we sample microstates with
probability proportional to $[G(E)]^{-1}$, where $G(E)$ is the
DOS, thus producing a flat energy histogram. The prerequisite for
the implementation of the method, is the DOS information of the
system, a problem that can now be handled in many adequate ways
via a number of interesting approaches proposed in the last two
decades. The most remarkable examples are the Lee
entropic~\cite{lee-93,lee-06}, the
multicanonical~\cite{berg-92,smith-95}, the broad
histogram~\cite{oliveira-96}, the transition
matrix~\cite{wang-99}, the WL~\cite{WL-01}, and the optimal
ensemble methods~\cite{trebst-04}.

In particular, the WL algorithm~\cite{WL-01} is one of the most
refreshing variations of the Monte Carlo simulation methods
introduced in the last years. The algorithm has already been
successfully used in many problems of statistical physics,
biophysics, and
others~\cite{douarche-03,troyer-03,fytas-06,schulz-05,reynal-05,jayasri-05,trebst-05,
rathore-02,shell-02,yamaguchi-01,carri-05,calvo-03,tsai-07,vorontsov-04,poulain-06,schulz-03}.
To apply the WL algorithm, an appropriate energy range  of
interest has to be identified and a WL random walk is performed in
this energy subspace. Trials from a spin state with energy $E_{i}$
to a spin state with energy $E_{f}$, using local spin flip
dynamics, are accepted according to the transition probability
\begin{equation}
\label{eq:1}p(E_{i}\rightarrow
E_{f})=\min\left[\frac{G(E_{i})}{G(E_{f})},1\right].
\end{equation}
During the WL process the DOS $G(E)$ is modified ($G(E)\rightarrow
f*G(E)$) after each spin flip trial by a modification factor
$f>1$. In the WL process ($j=1,2,\ldots,j_{f}$) successive
refinements of the DOS are achieved by decreasing the modification
factor $f_{j}$. Most implementations use an initial modification
factor $f_{j=1}=e\approx 2.71828\ldots$, a rule
$f_{j+1}=\sqrt{f_{j}}$, and a $5\%-10\%$ flatness criterion (on
the energy histogram) in order to move to the next refinement
level ($j\rightarrow j+1$)~\cite{WL-01}. The process is terminated
in a sufficiently high-level ($f\approx 1$, whereas the detailed
balanced condition limit is $f\rightarrow 1$).

In the last few years, there have been several papers dealing with
improvements and sophisticated implementations of the WL iterative
process~\cite{BP-07,lee-06,troyer-03,shell-02,poulain-06,dayal-04,alder-04,zhou-05}.
The present authors have introduced a dominant energy subspace
implementation of the above entropic methods, called critical
minimum energy subspace (CrMES) method~\cite{malakis-04}. This is
a method of a systematic restriction of the energy space, with
increasing lattice size, by which one can determine all
finite-size thermal anomalies of the system from the final
accurate DOS, and also other (magnetic) anomalies of the system by
accumulating appropriate histogram data in the final almost
entropic stage of the process. The (WL) random walk takes place in
a restricted energy subspace $(E_{1},E_{2})$ and this practice
produces an immense speed up, without introducing observable
errors. It has been shown that the method can be followed
successfully in pure systems undergoing second- or first-order
phase transitions~\cite{malakis-04,fytas-06} and also in more
complex systems with complicated free energy landscapes, such as
the 3D random field Ising model~\cite{fytas-08a}.

For the simulation of the random bond models considered in this
paper, we followed the general framework of the implementation of
the above described scheme. In particular, we followed most of the
details of the implementation applied recently to the 3D random
field Ising model and outlined for the random bond version of the
SAF model in our recently published Letter~\cite{fytas-08b}. In
these papers, a two stage strategy has been followed. In the first
stage, a multi-range (multi-R) WL method was applied, where the
total energy range was split in many subintervals~\cite{WL-01} and
the DOS's of these separate pieces were then joined at the end of
the process. The WL refinement levels used in this first multi-R
WL stage $(j=1,\ldots,j_{i})$, were as follows: $j_{i}=18$ for
$L<80$, $j_{i}=19$ for $80\leq L < 120$, and $j_{i}=20$ for $L\geq
120$. In the second stage of the simulation (WL refinement levels:
$j=j_{i}-j_{f}$), a more demanding multi-R - but with larger
energy pieces - or an one-range (one-R) approach was carried out.
The identification of the appropriate energy subspace
$(E_{1},E_{2})$ for the entropic sampling of each disorder
realization was carried out by applying our CrMES
restriction~\cite{malakis-04} and taking the union subspace at
both pseudocritical temperatures of the specific heat and
susceptibility. This union subspace, extended by $10\%$ from each
side, low- and high-energy side, is in most cases sufficient for
an accurate estimation of all finite-size anomalies. The
identification of the appropriate energy subspace was carried out
in the first multi-R WL stage, using originally a very wide energy
subspace. After the first identification, the same first stage
process ($j=1,\ldots,j_{i}$) was repeated several times, typically
$\sim 4-6$ times, in the new restricted energy subspace. From our
experience, this repeated application of this first stage multi-R
WL approach greatly improves accuracy, and then the resulting
accurate DOS is used again for a final and more accurate
redefinition of the subspace $(E_{1},E_{2})$, in which the final
entropic scheme (second stage) is applied. Thus, the final
entropic WL stage $(j=j_{i}-j_{f})$, where $j_{f}=j_{i}+4$ in all
cases, was carried out in this accurately defined subspace in an
one-R or in a multi-R approach. For the present models, it was
found that a final multi-R approach with large subranges (see
below) is in fact sufficiently accurate. Therefore, since the
multi-R of the original WL scheme improves efficiency, we applied,
for most of our simulations, this multi-R WL approach also in the
final entropic stage, using three times larger energy subintervals
than in the initial multi-R stage. The energy subintervals of the
first stage where chosen to correspond to rather large subspaces,
with their sizes depending on the disorder strength. Taking the
pure system as reference, these energy subintervals could be
chosen of the order of $50$ to $100$ energy levels, depending on
the lattice size. In the disorder case, the subinterval sizes are
multiplied by the factor induced due to the new multiplicity of
energy levels, giving for instance, a factor 4 for the disorder
strength $r=3/5=0.6$, where $r$ is the ratio of weak over strong
bond interactions (see also next Section).

As pointed out above, the need of using in the final stage the
described multi-R approach, instead of an one-R approach, is a
consequence of the slow convergence at the high WL levels. It is
possible to overcome this slow convergence by using a looser
flatness criterion or an alternative Lee entropic final stage, as
proposed in reference~\cite{lee-06} and applied by the present
authors~\cite{fytas-08a}. However, recently a different
alternative has been proposed by Belardinelli and Pereyra
(BP)~\cite{BP-07}, which is free of the application of the
energy-histogram flatness criterion. Following their proposal, one
is using, in the final stage, an almost continuously changing
modification factor adjusted according to the rule $\ln{f}\sim
t^{-1}$. Since $t$ is the Monte Carlo time, using a time-step
conveniently defined proportional to the size of the energy
subinterval, the efficiency of this scheme is independent of the
size of the subintervals and therefore the method provides the
same efficiency in both multi-R and one-R approaches. Furthermore,
from the tests performed by these authors, and also from our
comparative studies in the 2D pure Ising model (unpublished), the
error-behavior of this method seems superior to the original WL
process, improving to some extent the saturation-error problem of
the WL method. Accordingly, we have also applied this alternative
route for the final stage of our simulations using an one-R
approach. In particular, the disorder strength case $r=9/11=0.818$
of the random bond square SAF model, and all simulations
corresponding to the larger sizes $L=160$ and $L=200$ of the
random bond Ising model were carried out using this alternative.
From our comparative tests, we found that both approaches (the BP
and the multi-R WL approach) produced very accurate results, with
the BP approach giving superior estimates for the pseudocritical
temperatures of the models.

Both disordered models were simulated for two values of the
disorder strength $r$. For the random bond square SAF model we
chose the values $r=9/11=0.818$ and $r=3/5=0.6$, whereas for the
random bond Ising model the values $r=3/5=0.6$ and $r=1/7=0.142$
have been considered. Square lattices, using periodic boundary
conditions, with linear sizes $L$ in the range $L=20-120$ or
$L=20-200$ (disorder strength case $r=1/7=0.142$) were used. A
number of $200$, disorder realizations was generated and simulated
for each disorder case and lattice size. Even for the larger
lattice sizes the statistical errors of the WL method (WL-errors),
used for the estimation of thermal and magnetic properties of a
particular realization, were found to be much smaller than the
statistical errors of the disorder averaging, coming from the fact
that we have used a finite number of $200$ disorder realizations.
Therefore, the WL-errors are not shown in our graphs, whereas the
latter errors of finite disorder sampling are presented in our
figures as error bars. The mean values over disorder will be
denoted as $[\ldots]_{av}$, the corresponding maxima as
$[\ldots]^{\ast}_{av}$, and finally the individual maxima as
$[\ldots^{\ast}]_{av}$. Since in the fitting attempts of the
following Sections, we have used mainly data from the peaks of the
disorder averaged curves (i.e. $[C]^{\ast}_{av}$), their finite
disorder sampling errors are the relevant statistical errors to be
used in the fitting attempts. These errors have been estimated by
two similar methods, using groups of $25$ to $50$ realizations for
each lattice size and the jackknife method or a straightforward
variance calculation (blocking method)~\cite{newman-99}. The
jackknife method yielded some reasonably conservative errors,
about $10-20\%$ larger than the corresponding calculated standard
deviations, and are shown as error bars in our figures. Finally,
let us point out that in all cases studied, the sample-to-sample
fluctuations for the individual maxima are much large than the
corresponding finite disorder sampling errors.

\section{The random bond square SAF model: Competing interactions}
\label{sec:3}

In this Section we extend our investigation on the effects of
quenched bond randomness on the square Ising model with nearest-
($J_{nn}$) and next-nearest-neighbor ($J_{nnn}$) antiferromagnetic
interactions. In zero field, the pure square SAF model, is
governed by the Hamiltonian:
\begin{equation}
\label{eq:2}
\mathcal{H}_{p}=J_{nn}\sum_{<i,j>}S_{i}S_{j}+J_{nnn}\sum_{(i,j)}S_{i}S_{j},
\end{equation}
where here both nearest- ($J_{nn}$) and next-nearest-neighbor
($J_{nnn}$) interactions are assumed to be positive. It is
well-known that the model develops at low temperatures SAF order
for $R=J_{nn}/J_{nnn}>0.5$~\cite{swendsen-79,binder-80,oitmaa-87}
and by symmetry the critical behavior associated with the SAF
ordering is the same under $J_{nn}\rightarrow -J_{nn}$. For the
case $R=1$, that we deal with, the pure system undergoes a
second-order phase transition, in accordance with the commonly
accepted scenario for many years of a non-universal critical
behavior with exponents depending on the coupling ratio
$R$~\cite{swendsen-79,binder-80,landau-85,tanaka-92,minami-94}.
The recent numerical study of Malakis \etal~\cite{malakis-06} has
refined earlier estimates~\cite{binder-80,tanaka-92} for the
correlation length exponent $\nu$ and values very close to those
of the 2D three-state Potts model
$\nu_{p}$(Potts)$=5/6$~\cite{wu-82} were obtained. From the FSS of
the pseudocritical temperatures~\cite{malakis-06} it was found
that $\nu_{p}$(SAF;$R=1$)$=0.8330(30)$ and the subsequent study of
Monroe and Kim~\cite{monroe-07}, using the Fisher zeroes of the
partition function, yielded a quite matching estimate:
$\nu_{p}$(SAF;$R=1$)$=0.848(1)$. Furthermore, from the FSS of the
specific heat data an estimate for the ratio
$\alpha_{p}/\nu_{p}=0.412(5)$ was also found~\cite{malakis-06}.
Finally, from the magnetic data and in accordance with an earlier
conjecture of Binder and Landau~\cite{binder-80}, Malakis
\etal~\cite{malakis-06} found additional evidence of the weak
universality scenario~\cite{suzuki-74,gunton-75} and obtained the
values $\beta_{p}/\nu_{p}=0.125$ and $\gamma_{p}/\nu_{p}=1.75$.
\begin{figure}[ht]
\centerline{\includegraphics*[width=12 cm]{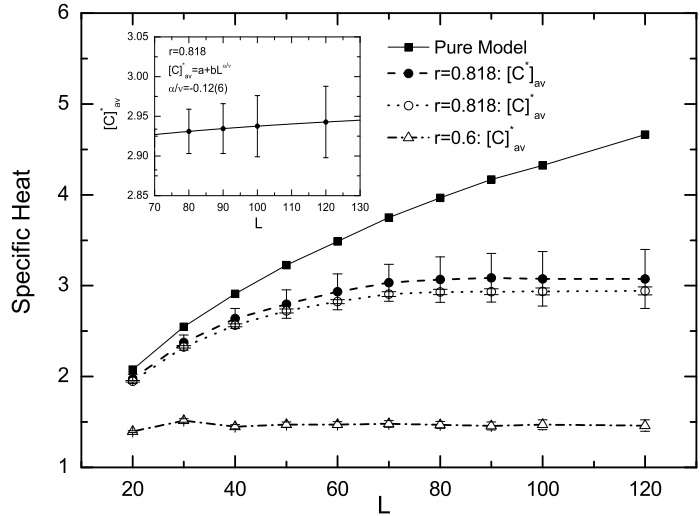}}
\caption{\label{fig:1}Size dependence of the maxima of the
specific heat for the pure (filled squares; data taken from
reference~\cite{malakis-06}) and the random bond (filled and open
circles for $r=0.818$ and open triangles for $r=0.6$) square SAF
model. The inset shows a power law fit for the case $r=0.818$ for
$L\geq 80$ giving a negative value for the exponent $\alpha/\nu$
of the order of $-0.12(6)$.}
\end{figure}
The values of the above three ratios of exponents satisfy the
Rushbrook relation, assuming that $\nu_{p}=0.8292$, which is very
close to the estimate obtained from the shift behavior of the SAF
$R=1$ model, thus providing self-consistency to the estimation
scheme. From these results, it is tempting to conjecture, as was
pointed out in reference~\cite{fytas-08b}, that the SAF model with
$R=1$ obeys the same thermal exponents with the 2D three-state
Potts model ($\nu_{p}=5/6=0.833\ldots$ and
$\alpha_{p}=1/3=0.333\ldots$~\cite{wu-82}), but the respective
values of the magnetic critical exponents are different
($\beta_{p}/\nu_{p}=2/15=0.133\ldots$ and
$\gamma_{p}/\nu_{p}=26/15=1.733\ldots$ for the 2D three-state
Potts model~\cite{wu-82}).

Considering now the random bond distribution~\cite{fytas-08b}
\begin{equation}
\label{eq:3}
P(J_{ij})=\frac{1}{2}[\delta(J_{ij}-J_{1})+\delta(J_{ij}-J_{2})];\;\;
\frac{J_{1}+J_{2}}{2}=1;\;\;r=\frac{J_{2}}{J_{1}},
\end{equation}
where the ratio $r$ stands for the disorder strength, the
Hamiltonian of equation~(\ref{eq:1}) is transformed into the
following disordered one
\begin{equation}
\label{eq:4}
\mathcal{H}=\sum_{<i,j>}J_{ij}S_{i}S_{j}+\sum_{(i,j)}J_{ij}S_{i}S_{j}.
\end{equation}
The critical behavior of the above defined random model for the
case $r=3/5=0.6$ has been outlined in reference~\cite{fytas-08b},
where apart from the verification of the weak universality
scenario, a strong saturating behavior of the specific heat has
been witnessed. Here, we extend our study to a weaker disorder
strength, namely the case $r=9/11=0.818$.
\begin{figure}[ht]
\centerline{\includegraphics*[width=14 cm]{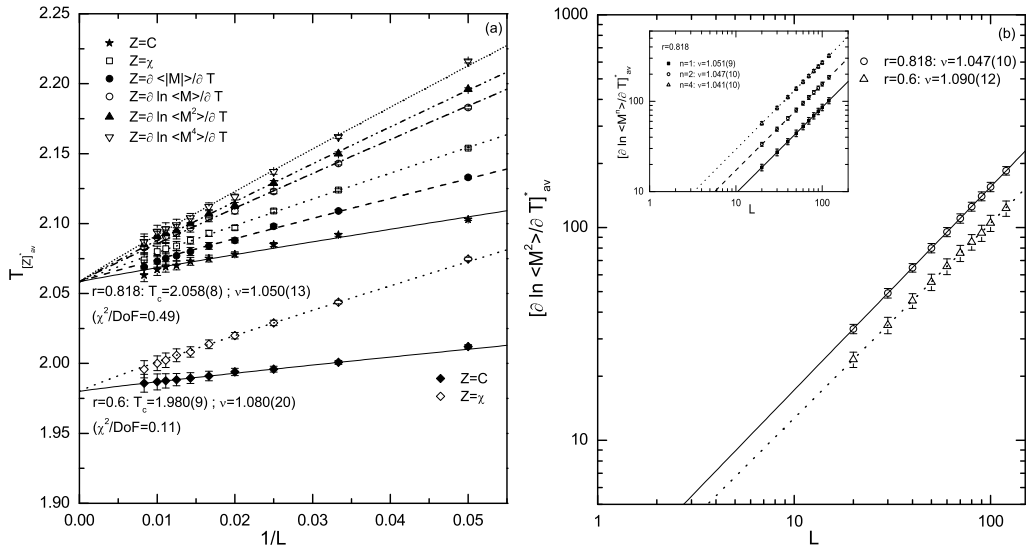}}
\caption{\label{fig:2} (a) Simultaneous fittings of several
pseudocritical temperatures of the random bond square SAF model
for the two values of $r$ considered: $r=0.818$ and $r=0.6$. (b)
Log-log plot of the maxima of the average second-order logarithmic
derivative of the order-parameter of the random bond square SAF
model also for $r=0.818$ and $r=0.6$. The inset shows a log-log
plot of the maxima of the average logarithmic derivatives of the
order-parameter of first-, second-, and fourth-order, for the case
$r=0.818$. Linear fits are applied for $L\geq 30$.}
\end{figure}

Let us start the presentation of our results with the most
striking effect of the bond randomness on the specific heat of the
square SAF model. In figure~\ref{fig:1} we contrast the size
dependence of the specific heat maxima of the pure (filled
squares) and the random bond model (filled and open circles for
$r=0.818$ and open triangles for the case $r=0.6$). For the case
$r=0.818$ we show two data set points for the specific heat,
corresponding to the two averaging processes discussed previously
in Section~\ref{sec:2}. Note that the error bars for the quantity
$[C^{\ast}]_{av}$ shown reflect the sample-to-sample fluctuations,
whereas all other error bars are statistical errors due to the
finite number of disorder realizations (jackknife errors discussed
in Section~\ref{sec:2}). The suppression of the specific heat
maxima is clear for both disorder strength values and of course it
is much stronger for the case $r=0.6$, for which a clear
saturation is observed even for the smaller sizes shown ($L=40$).
For the present value $r=0.818$ we show in the inset of
figure~\ref{fig:1} a power law fitting attempt of the form
$[C]^{\ast}_{av}\sim C_{\infty}+bL^{\alpha/\nu}$ for sizes $L\geq
80$ which gives a negative value for the exponent $\alpha/\nu$ of
the order of $-0.12(6)$. Notably, this value of $\alpha/\nu$ will
be shown to be compatible with the one obtained by an alternative
method via the Rushbrook relation.

\begin{figure}[ht]
\centerline{\includegraphics*[width=14 cm]{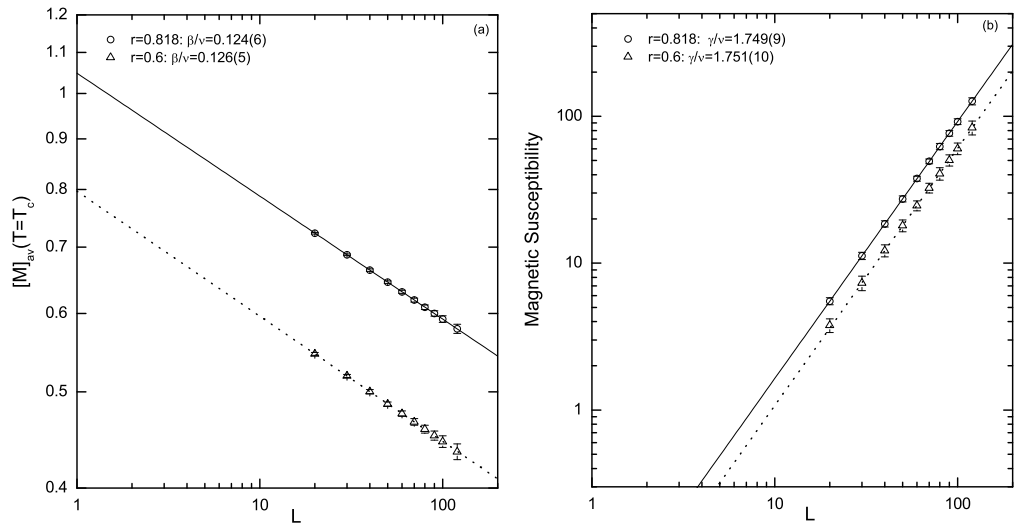}}
\caption{\label{fig:3} (a) Log-log plot of the size dependence of
the average magnetization of the random bond square SAF model at
the estimated corresponding critical temperatures for $r=0.818$
and $r=0.6$. (b) Log-lop plot of the size dependence of the maxima
of the average magnetic susceptibility of the random bond square
SAF model for $r=0.818$ and $r=0.6$. In both panels, the solid and
dotted lines are corresponding linear fits for $L\geq 30$.}
\end{figure}
In figure~\ref{fig:2}(a) we present the FSS behavior of several
pseudocritical temperatures of the model $T_{[Z]^{\ast}_{av}}$,
i.e. the temperatures corresponding to the maxima of several
average curves, such as the specific heat ($Z=C$), the magnetic
susceptibility ($Z=\chi$), the absolute order-parameter derivative
with respect to the temperature ($Z=\frac{\partial <|M|>}{\partial
T}$), and the logarithmic derivatives of several powers of the
order-parameter with respect to the temperature ($Z=\frac{\partial
\ln<M^{n}>}{\partial T}$, $n=1,2$ and $4$). For the case $r=0.818$
all the above $6$ mentioned pseudocritical temperatures are
presented and are compared to the case $r=0.6$~\cite{fytas-08b}.
The lines show two sets of simultaneous fitting attempts,
according to the shift relation
\begin{equation}
\label{eq:5} T_{[Z]^{\ast}_{av}}=T_{c}+bL^{-1/\nu},
\end{equation}
giving $T_{c}=2.058(8)$ for $r=0.818$ and $T_{c}=1.980(9)$ for
$r=0.6$, respectively, for the critical temperature of the
disordered model. Note that the corresponding critical temperature
of the pure system is $T_{c;p}=2.0823(17)$~\cite{malakis-06}. The
values for $\chi^{2}$/DoF of the fits shown depend on the method
used to evaluate the statistical errors (jackknife or simple
standard deviation errors) and in all cases studied in this paper
vary in the range $0.1-0.5$. A first estimation of the critical
exponent $\nu$ of the correlation length is obtained from the
above shift behavior and is $\nu=1.050(13)$ and $\nu=1.080(20)$
for $r=0.818$ and $r=0.6$ respectively, as illustrated in the
graph. An alternative estimation of the exponent $\nu$ is
attempted now from the FSS analysis of the logarithmic derivatives
of the order-parameter~\cite{order} with respect to the
temperature~\cite{chen-95,ferrenberg-91}
\begin{equation}
\label{eq:6} \frac{\partial \ln \langle M^{n}\rangle}{\partial
T}=\frac{\langle M^{n}E\rangle}{\langle M^{n}\rangle}-\langle
E\rangle,
\end{equation}
which scale as $L^{1/\nu}$ with the system size. In
figure~\ref{fig:2}(b) we consider in a log-log scale the size
dependence of the maxima of the average second-order logarithmic
derivative of the order-parameter for $r=0.818$ (open circles) and
$r=0.6$ (open triangles). The solid and dotted lines are
corresponding linear fits whose slopes provides respectively
estimates for $1/\nu$ and thus for the exponent $\nu$. It is clear
for the figure that the slopes of the lines are different and the
results we obtain from the linear fits are are $\nu=1.047(10)$ and
$\nu=1.090(12)$ for $r=0.818$ and $r=0.6$, respectively. In the
corresponding inset of figure~\ref{fig:2}(b) we present the first-
(filled squares), second- (open circles), and fourth-order (open
triangles) maxima of the average over the ensemble of realizations
logarithmic derivatives of the order-parameter for the case
$r=0.818$. The solid, dashed, and dotted lines shown are
corresponding linear fits whose slopes provide an average estimate
for the exponent $\nu$ of the order of $\nu=1.046(10)$. These
results for the exponent $\nu$ for both values of the disorder
strength $r$ compare favorably with the estimation of $\nu$ from
the above shift behavior shown in panel (a) of figure~\ref{fig:2}.
Thus, in comparison with its value of the pure model, the exponent
$\nu$ for the disordered model shows an increase of the order of
$20\%$ for the case $r=0.818$ and $30\%$ for the case $r=0.6$,
reflecting, in both cases, the strong influence of the disorder on
the thermal properties of the system. Noteworthy that, our
estimates for the exponent $\nu$ are in agreement with the
inequality $\nu\geq 2/D$ derived by Chayes \etal~\cite{chayes-86}
for disordered systems.

Turning now to the magnetic properties of the model we present in
figure~\ref{fig:3}(a) in a log-log scale the FSS behavior of the
average order-parameter for $r=0.818$ (open circles) at
$T_{c}=2.058$ and $r=0.6$ (open triangles) at $T_{c}=1.98$. The
straight lines show linear fits for $L\geq 30$ with a slope of
$0.124(6)$ for $r=0.818$ and $0.126(5)$ for $r=0.6$. Thus, these
two estimations for the ratio $\beta/\nu$ indicate that although
the exponent $\beta$ increases with disorder, the ratio
$\beta/\nu$ remains unchanged to its pure value, i.e.
$\beta/\nu=\beta_{p}/\nu_{p}=0.125$~\cite{fytas-08b,malakis-06}.
Correspondingly, we show in panel (b) of figure~\ref{fig:3} the
behavior of the magnetic susceptibility that provides estimates
for the ratio $\gamma/\nu$. The open circles refer to the case
$r=0.818$ and the open triangles to the case $r=0.6$. The solid
and dotted lines are linear fits giving the values
$\gamma/\nu=1.749(9)$ and $\gamma/\nu=1.751(10)$, respectively.
Thus, the ratio $\gamma/\nu$ maintains the value of the pure model
for both disorder strength values considered. The ratios
$\beta/\nu$ and $\gamma/\nu$ for the disordered square SAF model
appear to be the same with the corresponding ratios of the pure
square SAF model but different from those of the 2D three-state
Potts model. Therefore, our results reinforce both the weak
universality scenario for the pure SAF model, as first predicted
by Binder and Landau~\cite{binder-80}, as well as the generalized
statement of weak universality in the presence of bond randomness,
given by Kim~\cite{kim-96} and concerning also the 2D random bond
three-state Potts ferromagnet.

Finally, having estimated the ratios $\beta/\nu$, $\gamma/\nu$,
and the exponent $\nu$, we attempt to estimate the specific heat
exponent $\alpha$ using either the Rushbrook
($\alpha+2\beta+\gamma=2$) or equivalently, since
$2\beta/\nu+\gamma/\nu=2$, the hyperscaling ($2-\alpha=D \nu$)
relation. For the case $r=0.818$ we find a value
$\alpha/\nu=-0.09(4)$ which is in agreement with the value
$\alpha/\nu=-0.12(6)$ estimated from the fitting of the larger
specific heat data shown in the inset of figure~\ref{fig:1}. In
the case of a non-divergent specific heat it is quite interesting
to consider also, rather than the specific heat, the internal
energy scaling at the estimated critical temperature. This
provides an estimate for the exponent ratio $(\alpha-1)/\nu$,
which may be more precise~\cite{holm-94}. We finally carried out
such a fitting using the larger sizes ($L\geq 80$) on the values
of the critical energy (at $T_{c}=2.058$) and we found
$(\alpha-1)/\nu=-1.04(4)$. This result, when combined with the
estimate for $\nu$ from the shift behavior ($\nu=1.050(13)$) gives
a value $\alpha/\nu=-0.09(3)$, an intriguing coincidence, in full
agreement with the above value obtained for this ratio, via the
Rushbrook relation. For the case $r=0.6$~\cite{fytas-08b}, the
saturation effect is much more stronger and an even more negative
value for the specific heat exponent $\alpha$ is obtained
$\alpha=-0.17(4)$. From the above and in agreement with our
preliminary observation in reference~\cite{fytas-08b}, it turns
out that this strong saturating behavior of the specific heat is
completely different from the behavior of the 2D random bond
three-state Potts ferromagnet~\cite{kim-96,picco-96}. There, a
specific heat diverging behavior is obtained for disorder
strengths $r=0.9$, $0.5$, and $0.25$~\cite{kim-96} and an
increasing but progressively saturating behavior is obtained only
for the very strong disorder $r=0.1$\cite{picco-96}. This behavior
of the specific heat of the random bond square SAF model may
presumably be attributed to the competitive nature of
interactions, responsible for the observed sensitivity of the SAF
model to bond randomness.

\section{The marginal case of the 2D random bond Ising model}
\label{sec:4}

The Hamiltonian of the random bond version of the 2D Ising model
system is given by
\begin{equation}
\label{eq:7} \mathcal{H}=-\sum_{<i,j>}J_{ij}S_{i}S_{j},
\end{equation}
where again the implementation of the bond disorder follows the
binary distribution~(\ref{eq:3}) of ferromagnetic interaction
strengths. With this distribution the random Ising system exhibits
a unique advantage that its critical temperature $T_{c}$ is
exactly known~\cite{fisch-78,kinzel-81} as a function of the
disorder strength $r$ through
\begin{equation}
\label{eq:8} \sinh{(2J_{1}/T_{c})}\sinh{(2rJ_{1}/T_{c})}=1,
\end{equation}
where $r=J_{2}/J_{1}$ and $k_{B}=1$. This gives the opportunity of
carrying out the FSS analysis at the exact $T_{c}(r)$, a practice
that highly reduces statistical errors. Furthermore, one may check
for accuracy his numerical scheme by comparing the estimated
critical temperature (via the shift behavior) with the exact
result, as we have also done here.

At this is point, it is useful to briefly discuss the two main and
mutually excluded scenarios~\cite{MK-99} mentioned in the
introduction. The logarithmic corrections scenario is based on the
quantum field theory results of Dotsenko and
Dotsenko~\cite{DD-81}, and of Shalaev~\cite{Shalaev-84},
Shankar~\cite{Shankar-87}, and Ludwig~\cite{Ludwig-87}. According
to this scenario - supported theoretically for the case of weak
disorder - the presence of quenched disorder changes the critical
properties of the system only through a set of logarithmic
corrections to the pure system behavior. The specific heat $C$ is
expected to diverge on approach to the critical temperature
$T_{c}$ in a double logarithmic form: $C \propto \ln (\ln t)$,
where $t=|T-T_{c}|/T_{c}$ is the reduced critical temperature.
\begin{figure}[ht]
\centerline{\includegraphics*[width=16 cm]{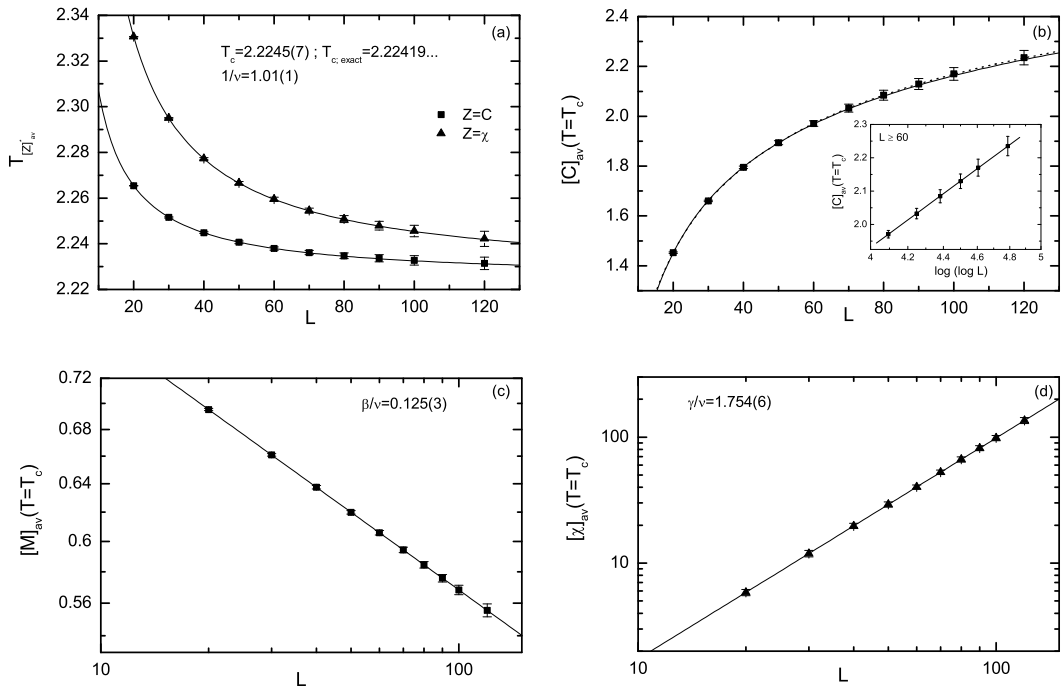}}
\caption{\label{fig:4} Critical behavior of the 2D random bond
Ising model for $r=3/5=0.6$. (a) Simultaneous fitting of two
pseudocritical temperatures defined in the text and estimation of
the correlation length exponent. (b) FSS behavior of the averaged
specific heat at $T_{c}(r=0.6)$. The solid and dotted lines are
corresponding double logarithmic and power law fits (see
equations~(\ref{eq:9}) and (\ref{eq:10}) in the text). The inset
shows the specific heat data as a function of the double logarithm
of the lattice size. A linear fit for $L\geq 60$ is applied. (c)
Log-log plot of the averaged magnetization at $T_{c}(r=0.6)$. (d)
Log-log plot of the averaged magnetic susceptibility at
$T_{c}(r=0.6)$. In both panels (c) and (d) a linear fit is
applied. Note that all fitting attempts shown have been performed
for $L\geq 20$.}
\end{figure}
On the other hand, according to the weak universality
scenario~\cite{KP-94,Kuhn-94,suzuki-74,gunton-75}, critical
quantities, such as the zero field susceptibility, magnetization,
and correlation length display power law singularities, with the
corresponding exponents $\gamma$, $\beta$, and $\nu$ changing
continuously with the disorder strength; however this variation is
such that the ratios $\gamma/\nu$ and $\beta/\nu$ remain constant
at the pure system's value. The specific heat of the disordered
system is, in this case, expected to saturate~\cite{KP-94}.

Two significantly different values of the disorder strength,
namely the cases $r=3/5=0.6$ and $r=1/7=0.142$ have been
investigated by our numerical scheme and our data will be
presented and analyzed below. Due to marginality, the case
$r=0.6$, that produced in the SAF model a $5\%$ temperature
decline, gives here, through equation~(\ref{eq:8}) a critical
temperature $T_{c}(r=0.6)=2.22419\ldots$, very close ($2\%$) to
the corresponding critical temperature of the pure system
($T_{c;p}=2.26918\ldots$). Thus, we may call the first case
($r=3/5=0.6$) a weak disorder and the second case ($r=1/7=0.142$)
a strong disorder case, actually producing a significant
temperature decline $T_{c}(r=0.142)=1.77910\ldots$. Our
simulations are extended to lattice sizes in the range $L=20-120$
for the weak disorder and in the range $L=20-200$ for the strong
disorder, in the hope that we will observe the true asymptotic
behavior of the model. For the calculation of the statistical
errors due to the finite number of simulated realizations, we have
followed again the practice outlined in Section~\ref{sec:2}. The
values of $\chi^{2}$/DoF of all the fits shown below are again in
the range $0.1-0.5$.

We start the presentation of our results with the case $r=3/5=0.6$
and our fitting attempts are summarized in figure~\ref{fig:4}.
Figure~\ref{fig:4}(a) shows a simultaneous fitting - including all
data points - using the shift behavior~(\ref{eq:5}) for the
estimation of the critical temperature and the correlation length
exponent. Now, we have used the pseudocritical temperatures
$T_{[Z]^{\ast}_{av}}$ of the specific heat ($Z=C$) and magnetic
susceptibility ($Z=\chi$). As shown in the panel, the estimated
value for the critical temperature is $T_{c}=2.2245(7)$ in
excellent agreement with the exact value. Respectively, the
estimation of the inverse of the correlation length exponent is
$1/\nu=1.01(1)$, which provides a value for $\nu$ of the order
$\nu=0.99(1)$, very close to the value $\nu=1$ of the pure model.

In panel (b) of figure~\ref{fig:4} we illustrate the data of the
averaged specific heat at the exact critical temperature. The
solid and dotted lines are corresponding fits of the form
\begin{equation}
\label{eq:9} [C]_{av}(T=T_{c}) \sim C_{1}+C_{2}\ln{(\ln L)}
\end{equation}
and
\begin{equation}
\label{eq:10} [C]_{av}(T=T_{c}) \sim C_{\infty}+b L^{\alpha/\nu}.
\end{equation}
As it clear from this graph, one may not easily discern between
the two lines, although a more careful analysis indicates that the
double logarithmic function is more stable. We have performed both
kinds of fits for three sets of data points $L_{min}-L_{max}$. We
fixed $L_{max}=120$ and varied $L_{min}$ as follows: $L_{min}=20$,
$L_{min}=50$, and $L_{min}=80$. We have observed that although the
values of $\chi^{2}$/DoF are comparable between the two
functions~(\ref{eq:9}) and (\ref{eq:10}), the coefficient $C_{2}$
of equation~(\ref{eq:9}) seems to be stable ($C_{2}\simeq
1.67(3)$), whereas the exponent $\alpha/\nu$ of
equation~(\ref{eq:10}) fluctuates, with increasing $L_{min}$, in
the range $-0.3$ to $-0.2$:
$\frac{\alpha}{\nu}(L_{min}=20)=-0.24(2)$,
$\frac{\alpha}{\nu}(L_{min}=50)=-0.21(4)$, and
$\frac{\alpha}{\nu}(L_{min}=80)=-0.30(3)$. Although for this case
we can not conclusively discriminate between the two alternatives,
the inset in panel (b) shows that the specific heat data fits well
to the double logarithm for lattice sizes $L\geq 60$. Finally, in
panels (c) and (d) we plot the average magnetization and the
average magnetic susceptibility at the critical temperature
respectively, as a function of the lattice size $L$ in a log-log
scale. In both cases a linear fit is applied for $L\geq 20$ giving
the values $\beta/\nu=0.125(3)$ and $\gamma/\nu=1.754(6)$, i.e.
the ratios of the magnetic exponents $\beta/\nu$ and $\gamma/\nu$
maintain the values $\beta_{p}/\nu_{p}=0.125$ and
$\gamma_{p}/\nu_{p}=1.75$ of the pure model.
\begin{figure}[ht]
\centerline{\includegraphics*[width=16 cm]{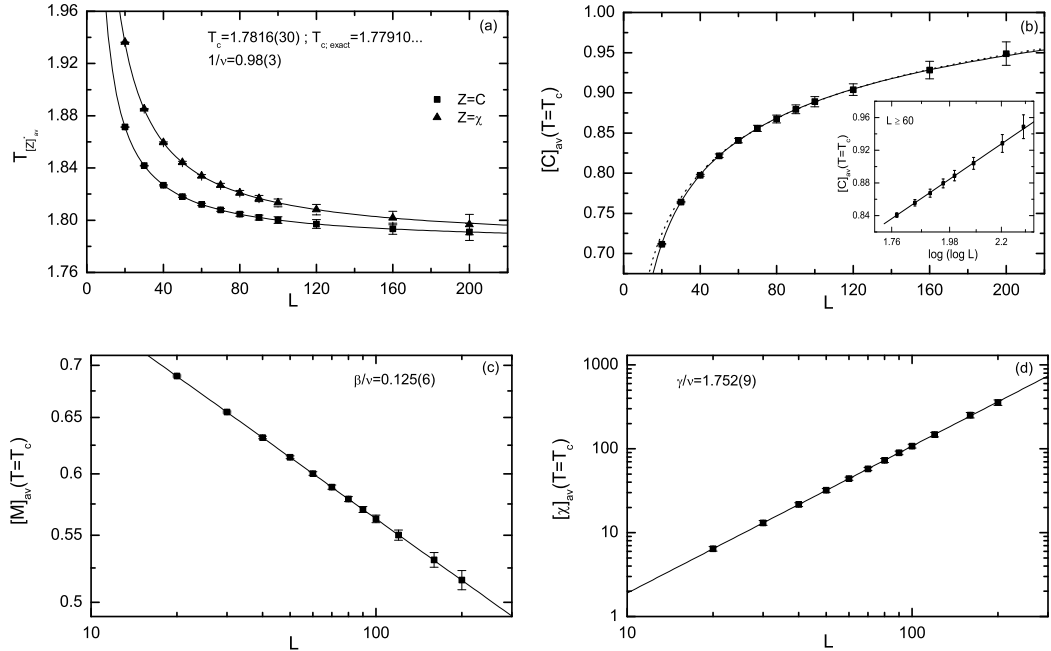}}
\caption{\label{fig:5} The same with figure~\ref{fig:4} for
$r=1/7=0.142$. Now data for lattice sizes up to $L=200$ are
presented. All fitting attempts shown are performed for $L\geq
60$.}
\end{figure}

We proceed to present our results for the strong disorder strength
$r=1/7=0.142$. As was already stated above, for this value of $r$
we have extended our simulations up to sizes $L=200$.
Figure~\ref{fig:5} summarizes now the critical behavior for the
case $r=0.142$ and, in particular, in figure~\ref{fig:5}(a) we
estimate by a simultaneous fitting attempt for the larger lattices
($L\geq 60$) the critical temperature and the correlation length
exponent. The estimated value for the critical temperature is
$T_{c}=1.7816(30)$ in good agreement with the exact value. The
production of the above accurate estimates for the critical
temperatures, even in this strong disorder case, constitutes a
concrete reliability test, in favor of the accuracy of our
numerical scheme. The estimation of the inverse of the correlation
length exponent is $1/\nu=0.98(3)$, providing a value for $\nu$ of
the order $\nu=1.02(3)$, which within error bars agrees with the
pure model's correlation length exponent value. In panel (b) of
figure~\ref{fig:5} we plot the data of the averaged specific heat
at the exact critical temperature. Again, the solid and dotted
lines are corresponding fits of the forms~(\ref{eq:9}) and
(\ref{eq:10}) for the larger sizes ($L\geq 60$). Our analysis for
the quality of the fits for the three sets of data points
$L_{min}-L_{max}$, with $L_{max}=200$ and $L_{min}=20$, $50$, and
$80$ indicated a good trend for the values of $\chi^{2}$/DoF for
the double logarithmic fits~(\ref{eq:9}) in the range: $0.1-0.3$.
Further reliability in favor of the logarithmic corrections
scenario is provided by the stability of the coefficient $C_{2}$:
$C_{2}(L_{min}=20)=0.412(4)$, $C_{2}(L_{min}=50)=0.410(6)$,
$C_{2}(L_{min}=80)=0.418(7)$, whereas the value of the exponent
$\alpha/\nu$ of the power law~(\ref{eq:10}) approaches zero with
increasing $L_{min}$: $\frac{\alpha}{\nu}(L_{min}=20)=-0.27(4)$,
$\frac{\alpha}{\nu}(L_{min}=50)=-0.20(6)$, and
$\frac{\alpha}{\nu}(L_{min}=80)=-0.02(3)$. From the above, we may
conclude that our numerical data are more properly described, at
least for the larger lattice sizes studied, by the double
logarithmic form~(\ref{eq:9}). The inset in panel (b) of
figure~\ref{fig:5} shows again the specific heat data as a
function of the double logarithm of the lattice size for $L\geq
60$. The solid line shown is an excellent linear fit. It should be
noted that, the above power law behavior ($\alpha/\nu\rightarrow
0_{-}$) is in full agreement with the earlier observation of Wang
\etal~\cite{wang-90} in the strong disorder regime ($r=1/4$ and
$r=1/10$)~\cite{wang-90}). Note also that, analogous results have
been presented in figure 2 of reference~\cite{BFMMPR-97} for the
site diluted Ising model, where sizes up to $L=256$ have been
considered. Finally, in panels (c) and (d) we plot the average
magnetization and the average magnetic susceptibility at the
critical temperature respectively, as a function of the lattice
size $L$ in a log-log scale. In both cases, the solid lines shown
are linear fits for $L\geq 60$ giving again within error bars the
values of the pure model, $\beta/\nu=0.125(6)$ and
$\gamma/\nu=1.752(9)$, respectively.

Summarizing, we may point out that the difficulties observed in
the case of weak disorder in discriminating between the two
scenarios have been surpassed by considering the strong disorder
case and larger lattice sizes. In particular, we take as evidence
in favor of the picture emerged from the theoretical work of
Dotsenko and Dotsenko~\cite{DD-81} and the improved versions of
Shalaev~\cite{Shalaev-84}, Shankar~\cite{Shankar-87}, and
Ludwig~\cite{Ludwig-87}, the stability of our fitting attempts
using the double logarithmic law. Similar conclusions have also
been reported by the extensive Monte Carlo studies of Wang
\etal~\cite{wang-90} for the random bond model and also of Selke
\etal~\cite{SSV-98}, Ballesteros \etal~\cite{BFMMPR-97}, and
Tomita and Okabe~\cite{TO-01} for the site diluted model, by the
transfer matrix approach of Ar\~ao Reis \etal~\cite{AQS-96,AQS-97}
for the random bond case, and very recently by Kenna and
Ruiz-Lorenzo~\cite{kenna-08} via an alternative approach that
involves the density of Lee-Yang zeros of the site diluted model.

\section{Conclusions}
\label{sec:5}

The effects induced by the presence of quenched bond randomness on
the critical behavior of two Ising spin models in 2D have been
investigated by a sophisticated entropic scheme based on the
Wang-Landau algorithm. For the random bond square SAF model we
have extracted accurate estimates for all critical exponents and
two values of the disorder strength. These values verify
hyperscaling, satisfy the Chayes \etal
inequality~\cite{chayes-86}, and obey very well the weak
universality scenario for disordered systems~\cite{kim-96}.
Furthermore, the strong saturating behavior of the specific heat
clearly distinguishes this case of competing interactions from
other 2D random bond systems studied previously. For the marginal
case of the random bond Ising model, our findings favor the
well-known double logarithmic scaling scenario and suggest that
the pure system behavior, $\nu=1$, is recovered in the asymptotic
limit. Here, the estimated critical temperatures, in both cases of
disorder, are in excellent agreement with the exact values
obtained by duality reflecting the accuracy of the implemented
entropic scheme. Encouraged by this latter observation, we are
currently carrying out a similar study of bond disorder effects in
2D systems undergoing first-order phase transitions.

\ack{We thank V Mart\'{i}n-Mayor and W Selke for useful e-mail
correspondence. Research supported by the special Account for
Research Grants of the University of Athens under Grant Nos.
70/4/4071 and 70/4/4096. N G Fytas acknowledges financial support
by the Alexander S. Onassis Public Benefit Foundation.}


\section*{References}

\begin{thebibliography}{170}

\bibitem{imry-79} Imry Y and Wortis M, 1979 \emph{Phys. Rev. B} {\bf 19} 3580

\bibitem{aizenman-89} Aizenman M and Wehr J, 1989 \emph{Phys. Rev. Lett.} {\bf 62}
2503; erratum 1990 {\bf 64} 1311

\bibitem{hui-89} Hui K and Berker A N, 1989 \emph{Phys. Rev. Lett.} {\bf 62}
2507; erratum 1989 {\bf 63} 2433; \\
Berker A N, 1990 \emph{Phys. Rev. B} {\bf 42}, 8640

\bibitem{chen-95} Chen S, Ferrenberg A M and Landau D P, 1995 \emph{Phys. Rev. E} {\bf 52} 1377

\bibitem{chatelain-01} Chatelain C, Berche B, Janke W and Berche P E, 2001 \emph{Phys. Rev. E} {\bf 64} 036120

\bibitem{fernandez-08} Fern\'{a}ndez L A, Gordillo-Guerrero A,
Mart\'{i}n-Mayor V and Ruiz-Lorenzo J J, 2008 \emph{Phys. Rev.
Lett.} {\bf 100} 057201

\bibitem{harris-74} Harris A B, 1974 \emph{J. Phys. C} {\bf 7} 1671

\bibitem{chayes-86} Chayes J T, Chayes L, Fisher D S and Spencer T, 1986 \emph{Phys. Rev. Lett.} {\bf 57} 2999

\bibitem{dotsenko-95} Dotsenko V, Picco M and Pujol P, 1995 \emph{Nucl. Phys. B} {\bf 455} 701

\bibitem{cardy-96} Cardy J, 1996 \emph{J. Phys. A} {\bf 29} 1897

\bibitem{jacobsen-98} Jacobsen J L and Cardy J, 1998 \emph{Nucl. Phys. B} {\bf 515} 701

\bibitem{olson-00} Olson T and Young A P, 2000 \emph{Phys. Rev. B} {\bf 61} 12467

\bibitem{chatelain-00} Chatelain C and Berche B, 2000 \emph{Nucl. Phys. B} {\bf 572} 626

\bibitem{MK-99} Mazzeo G and K\"uhn R, 1999 \emph{Phys. Rev. E} {\bf 60} 3823

\bibitem{fytas-08a} Fytas N G, Malakis A and Eftaxias K, 2008 \emph{J. Stat. Mech.} P03015

\bibitem{fytas-08b} Fytas N G, Malakis A and Georgiou I, 2008 \emph{J. Stat. Mech.} L07001

\bibitem{butera-08} Butera P and Pernici N, 2008 \emph{Phys. Rev. B} {\bf 78} 054405

\bibitem{DD-81} Dotsenko V S and Dotsenko V S, 1981 \emph{Sov. Phys. JETP Lett.} {\bf 33} 37; \\
Dotsenko V S and Dotsenko V S, 1983 \emph{Adv. Phys.} {\bf 32} 129

\bibitem{Shalaev-84} Shalaev B N, 1984 \emph{Sov. Phys. Solid State} {\bf 26} 1811

\bibitem{Cardy-86} Cardy J L, 1986 \emph{J. Phys. A} {\bf 19}, L193

\bibitem{Shankar-87} Shankar R, 1987 \emph{Phys. Rev. Lett.} {\bf 58} 2466; \\
Ludwig A W W, 1988 \emph{Phys. Rev. Lett.} {\bf 61} 2388; \\
Ceccatto H A and Naon C, 1988 \emph{Phys. Rev. Lett.} {\bf 61}
2389

\bibitem{Ludwig-87} Ludwig A W W, 1987 \emph{Nucl. Phys. B} {\bf 285} 97

\bibitem{LC-87} Ludwig A W W and Cardy J L, 1987 \emph{Nucl Phys. B} {\bf 285} 687

\bibitem{Mayer-89} Mayer I O, 1989 \emph{J. Phys. A} {\bf 22} 2815

\bibitem{wang-90} Wang J -S, Selke W, Dotsenko V S and Andreichenko V B, 1990 \emph{Physica A} {\bf 164} 221

\bibitem{Ludwig-90} Ludwig A W W, 1990 \emph{Nucl. Phys. B} {\bf 330} 639

\bibitem{Ziegler-90} Ziegler K, 1990 \emph{Nucl. Phys. B} {\bf 344} 499

\bibitem{WSDA-90} Wang J -S, Selke W, Dotsenko V S and Andreichenko V B, 1990 \emph{Europhys. Lett.} {\bf 11} 301

\bibitem{Heuer-92} Heuer H, 1992 \emph{Phys. Rev. B} {\bf 45} 5691

\bibitem{Shalaev-94} Shalaev B N, 1994 \emph{Phys. Rep.} {\bf 237} 129

\bibitem{KP-94} Kim J -K and Patrascioiu A, 1994 \emph{Phys. Rev. Lett.} {\bf 72} 2785; \\
Kim J -K and Patrascioiu A, 1994 \emph{Phys. Rev. B} {\bf 49}
15764; \\ Selke W, 1994 \emph{Phys. Rev. Lett.} {\bf 73} 3487; \\
Ziegler K, 1994 \emph{Phys. Rev. Lett.} {\bf 73} 3488; \\ Kim J -K
and Patrascioiu A, 1994 \emph{Phys. Rev. Lett.} {\bf 73} 3489; \\
Kim J -K, 2000 \emph{Phys. Rev. B} {\bf 61} 1246

\bibitem{Kuhn-94} K\"uhn R, 1994 \emph{Phys. Rev. Lett.} {\bf 73} 2268

\bibitem{QS-94} de Queiroz S L A and Stinchcombe R B, 1994 \emph{Phys. Rev. B} {\bf 50} 9976

\bibitem{TS-94} Talapov A L and Shchur L N, 1994 \emph{Europhys. Lett.} {\bf 27}
193; \\ Talapov A L and Shchur L N, 1994 \emph{J. Phys. C.} {\bf
6} 8295

\bibitem{MS-95} Mussardo G and Simonetti P, 1995 \emph{Phys. Lett. B} {\bf 351} 515

\bibitem{AQS-96} Ar\~ao Reis F D A, de Queiroz S L A and dos Santos R R, 1996 \emph{Phys. Rev. B} {\bf 54} R9616

\bibitem{JS-96} Jug G and Shalaev B N, 1996 \emph{Phys. Rev. B} {\bf 54} 3442

\bibitem{CHMP-97} Cabra D C, Honecker A, Mussardo G and Pujol P, 1997 \emph{J. Phys. A} {\bf 30} 8415

\bibitem{BFMMPR-97} Ballesteros H G, Fern\'andez L A, Mart\'in-Mayor V, Mu\~noz Sudupe A,
Parisi G and Ruiz-Lorenzo J J, 1997 \emph{J. Phys. A} {\bf 30}
8379

\bibitem{AQS-97} Ar\~ao Reis F D A, de Queiroz S L A and dos Santos R R, 1997 \emph{Phys. Rev. B} {\bf 56} 6013

\bibitem{SSLI-97} Selke W, Szalma F, Lajko P and Igloi F, 1997 \emph{J. Stat. Phys.} {\bf 89} 1079

\bibitem{RAJ-98} Roder A, Adler J and Janke W, 1998 \emph{Phys. Rev. Lett.} {\bf 80} 4697; \\
Roder A, Adler J and Janke W, 1999 \emph{Physica A} {\bf 265} 28

\bibitem{SSV-98} Selke W, Shchur L N and Vasilyev O A, 1998 \emph{Physica A} {\bf 259} 388

\bibitem{AQS-99} Ar\~ao Reis F D A, de Queiroz S L A and dos Santos R R, 1999 \emph{Phys. Rev. B} {\bf 60} 6740

\bibitem{LSZ-01} Luo H J, Sch\"ulke L and Zheng B, 2001 \emph{Phys. Rev. E} {\bf 64} 036123

\bibitem{Nobre-01} Nobre F D, 2001 \emph{Phys. Rev. E} {\bf 64} 046108

\bibitem{SV-01} Shchur L N and Vasilyev O A, 2001 \emph{Phys. Rev. E} {\bf 65} 016107

\bibitem{TO-01} Tomita Y and Okabe Y, 2001 \emph{Phys. Rev. E} {\bf 64} 036114

\bibitem{MC-02} Merz F and Chalker J T, 2002 \emph{Phys. Rev. B} {\bf 65} 054425

\bibitem{COPS-04} Calabrese P, Orlov E V, Pakhnin V and Sokolov A I, 2004 \emph{Phys. Rev. B} {\bf 70} 094425

\bibitem{Queiroz-06} de Queiroz S L A, 2006 \emph{Phys. Rev. B} {\bf 73} 064410

\bibitem{LQ-06} Lessa J C and de Queiroz S L A, 2006 \emph{Phys. Rev. E} {\bf 74} 021114

\bibitem{PHP-06} Picco M, Honecker A and Pujol P, 2006 \emph{J. Stat. Mech.} P09006

\bibitem{kenna-06} Kenna R, Johnston D A and Janke W, 2006 \emph{Phys. Rev. Lett.} {\bf 96}
115701; \\ Kenna R, Johnston D A and Janke W, 2006 \emph{Phys.
Rev. Lett.} {\bf 97} 155702

\bibitem{MP-07} Martins P H L and Plascak J A, 2007 \emph{Phys. Rev. E} {\bf 76} 012102

\bibitem{hasenbusch-08} Hasenbusch M, Toldin F P, Pelissetto A and Vicari E, 2008 \emph{Phys. Rev. E} {\bf 78} 011110

\bibitem{hadjiagapiou-08} Hadjiagapiou I A, Malakis A and Martinos S S, 2008 \emph{Physica A} {\bf 387} 2256

\bibitem{kenna-08} Kenna R and Ruiz-Lorenzo J J, 2008 \emph{Phys. Rev. E} {\bf 78} 031134

\bibitem{suzuki-74} Suzuki M, 1974 \emph{Prog. Theor. Phys.} {\bf 51} 1992

\bibitem{gunton-75} Gunton J D and Niemeijer T, 1975 \emph{Phys. Rev. B} {\bf 11} 567

\bibitem{WL-01} Wang F and Landau D P, 2001 \emph{Phys. Rev. Lett.} {\bf 86} 2050; \\
Wang F and Landau D P, 2001 \emph{Phys. Rev. E} {\bf 64} 056101

\bibitem{malakis-04} Malakis A, Peratzakis A and Fytas N G, 2004 \emph{Phys. Rev. E} {\bf 70}
066128;\\ Malakis A, Martinos S S, Hadjiagapiou I A, Fytas N G and
Kalozoumis P, 2005 \emph{Phys. Rev. E} {\bf 72} 066120

\bibitem{BP-07} Belardinelli R E and Pereyra V D, 2007 \emph{Phys. Rev. E} {\bf 75}
046701; \\ Belardinelli R E and Pereyra V D, 2007 \emph{J. Chem.
Phys.} {\bf 127} 184105

\bibitem{metropolis-53} Metropolis N, Rosenbluth A W, Rosenbluth M N and Teller A H, 1953 \emph{J. Chem. Phys.} {\bf 21} 1087

\bibitem{bortz-75} Bortz A B, Kalos M H and Lebowitz J L, 1975 \emph{J. Comput. Phys.} {\bf 17} 10

\bibitem{binder-97} Binder K, 1997 \emph{Rep. Prog. Phys.} {\bf 60} 487

\bibitem{newman-99} Newman M E J and Barkema G T, 1999 \textit{Monte Carlo Methods in
Statistical Physics}, Clarendon Press, Oxford

\bibitem{landau-00} Landau D P and Binder K, 2000 \textit{A Guide to Monte Carlo Simulations in
Statistical Physics}, Cambridge University Press, Cambridge

\bibitem{lee-93} Lee J, 1993 \emph{Phys. Rev. Lett.} {\bf 71} 211

\bibitem{lee-06} Lee H K, Okabe Y and Landau D P, 2006 \emph{Comput. Phys. Commun.} {\bf 175} 36

\bibitem{oliveira-96} de Oliveira P M C, Penna T J P and Herrmann H J, 1996 \emph{Braz. J. Phys.} {\bf 26} 677

\bibitem{wang-99} Wang J -S, Tay T K and Swendsen R H, 1999 \emph{Phys. Rev. Lett.} {\bf 82}
476;\\ Wang J -S and Swendsen R H, 2002 \emph{J. Stat. Phys.} {\bf
106} 245

\bibitem{berg-92} Berg B A and Neuhaus T, 1991 \emph{Phys. Lett. B} {\bf 276} 249;\\
Berg B A and Neuhaus T, 1992 \emph{Phys. Rev. Lett.} {\bf 68} 9

\bibitem{smith-95} Smith G R and Bruce A D, 1995 \emph{J. Phys. A} {\bf 28} 6623

\bibitem{torrie-97} Torrie G M and Valleau J -P, 1997 \emph{J. Comput. Phys.} {\bf 23} 187

\bibitem{swendsen-86} Swendsen R H and Wang J -S, 1986 \emph{Phys. Rev. Lett.} {\bf 57} 2607

\bibitem{geyer-91} Geyer C J, 1991 \textit{Computing Science and Statistics:
Proceedings of the 23rd Symposium on the interface}, ed. E.K.
Keramidas, Interface Foundation, Fairfax Station, New York, p.
156.

\bibitem{marinari-92} Marinari E and Parisi G, 1992 \emph{Europhys. Lett.} {\bf 19} 451

\bibitem{lyubartsev-92} Lyubartsev A P, Martsinovskii A A, Shevkunov S V and Vorontsov-Velyaminov P N, 1992 \emph{J. Chem. Phys.} {\bf 96} 1776

\bibitem{hukushima-96} Hukushima K and Nemoto K, 1996 \emph{J. Phys. Soc. Jpn.} {\bf 65} 1604

\bibitem{marinari-98} Marinari E, Parisi G, Ruiz-Lorenzo J, 1998,
\textit{Spin Glasses and Random Fields}, ed. A.P. Young,
Directions in Condensed Matter Physics, World Scientific,
Singapore, Vol. 12

\bibitem{trebst-04} Trebst S, Huse D A and Troyer M, 2004 \emph{Phys. Rev. E} {\bf 70} 046701

\bibitem{douarche-03} Douarche N, Calvo F, Pastor G M and Jensen P J, 2003 \emph{Eur. Phys. J. D} {\bf 24} 77

\bibitem{troyer-03} Troyer M, Wessel S and Alet F, 2003 \emph{Phys. Rev. Lett.} {\bf 90} 120201

\bibitem{fytas-06} Malakis A and Fytas N G, 2006 \emph{Phys. Rev. E} {\bf 73} 056114;\\
Malakis A and Fytas N G, 2006 \emph{Phys. Rev. E} {\bf 73} 016109
;\\ Malakis A and Fytas N G, 2006 \emph{Eur. Phys. J. B} {\bf 51}
257 ;\\ Malakis A, Fytas N G and Kalozoumis P, 2007 \emph{Physica A} {\bf 383} 351;\\
Fytas N G and Malakis A, 2008 \emph{Eur. Phys. J. B} {\bf 61} 111

\bibitem{schulz-05} Schulz B J, Binder K and M\"{u}ller M, 2005 \emph{Phys. Rev. E} {\bf 71} 046705

\bibitem{reynal-05} Reynal S and Diep H T, 2005 \emph{Phys. Rev. E} {\bf 72} 056710

\bibitem{jayasri-05} Jayasri D, Sastry V S S and Murthy K P N, 2005 \emph{Phys. Rev. E} {\bf 72} 036702

\bibitem{trebst-05} Trebst S, Gull E and Troyer M, 2005 \emph{J. Chem. Phys.} {\bf 123} 204501

\bibitem{rathore-02} Rathore N and de Pablo J J, 2002 \emph{J. Chem. Phys.} {\bf 116}
7225;\\ Rathore N, Knotts T A and de Pablo J J, 2003 \emph{J.
Chem. Phys.} {\bf 118} 4285;\\ Rathore N, Yan G and de Pablo J J,
2004 \emph{J. Chem. Phys.} {\bf 120} 5781;\\ Yan Q, Faller R and
de Pablo J J, 2002 \emph{J. Chem. Phys.} {\bf 116} 8745

\bibitem{shell-02} Shell M S, Debenedetti P G and Panagiotopoulos A Z, 2002 \emph{Phys. Rev. E} {\bf 66} 056703

\bibitem{yamaguchi-01} Yamaguchi C and Okabe Y, 2001 \emph{J. Phys. A} {\bf 34}
8781;\\ Okabe Y, Tomita Y and Yamaguchi C, 2002 \emph{Comput.
Phys. Commun.} {\bf 146} 63

\bibitem{carri-05} Varshney V and Carri G A,
2005 \emph{Phys. Rev. Lett.} {\bf 95} 168304;\\ Carri G A, Batman
R, Varshney V and Dirama T E, 2006 \emph{Polymer} {\bf 46} 3809

\bibitem{calvo-03} Calvo F, 2002 \emph{Mol. Phys.} {\bf 100} 3421;\\ Calvo F and Parneix
P, 2003 \emph{J. Chem. Phys.} {\bf 119} 256

\bibitem{tsai-07} Tsai S -H, Wang F and Landau D P, 2007 \emph{Phys. Rev. E} {\bf 75} 061108;\\
Tsai S -H, F Wang and Landau D P, 2008 \emph{Braz. J. Phys.} {\bf
38} 6;\\ Seaton D T, Mitchell S J and Landau D P, 2008 \emph{Braz.
J. Phys.} {\bf 38} 48;\\ Mitchell S J, Luiz Pereira F C and Landau
D P, 2008 \emph{Braz. J. Phys.} {\bf 38} 1

\bibitem{vorontsov-04} Vorontsov-Velyaminov P N, Volkov N A and Yurchenko A
A, 2004 \emph{J. Phys. A} {\bf 37} 1573;\\ Volkov N A,
Vorontsov-Velyaminov P N and Lyubartsev A P, 2007 \emph{Phys. Rev.
E} {\bf 75} 016705

\bibitem{poulain-06} Poulain P, Calvo F, Antoine R, Broyer M and Dugourd
P, 2006 \emph{Phys. Rev. E} {\bf 73} 056704

\bibitem{schulz-03} Schulz B J, Binder K, M\"{u}ller M and Landau D P, 2003 \emph{Phys. Rev. E} {\bf 67} 067102

\bibitem{dayal-04} Dayal P, Trebst S, Wessel S, W\"{u}rtz D, Troyer M, Sabhapandit S and Coppersmith S N 2004
\emph{Phys. Rev. Lett.} {\bf 92} 097201

\bibitem{alder-04} Alder S, Trebst S, Hartmann A K and Troyer M 2004 \emph{J. Stat. Mech.} P07008

\bibitem{zhou-05} Zhou C and Bhatt R N 2005 \emph{Phys. Rev. E} {\bf 72} 025701(R);\\
Zhou C, Schulthess T C, Torbr\"{u}gge S and Landau D P 2006
\emph{Phys. Rev. Lett.} {\bf 96} 120201

\bibitem{swendsen-79} Swendsen R H and Krinsky S, 1979 \emph{Phys. Rev. Lett.} {\bf 43} 177

\bibitem{binder-80} Binder K and Landau D P, 1980 \emph{Phys. Rev. B} {\bf 21} 1941

\bibitem{oitmaa-87} Oitmaa J and Velgakis M J, 1987 \emph{J. Phys. A} {\bf 20} 1269

\bibitem{landau-85} Landau D P and Binder K, 1985 \emph{Phys. Rev. B} {\bf 31} 5946

\bibitem{tanaka-92} Tanaka K, Horiguchi T and Morita T, 1992 \emph{Phys. Lett. A} {\bf 165} 266

\bibitem{minami-94} Minami K and Suzuki M, 1994 \emph{J. Phys. A} {\bf 27} 7301

\bibitem{malakis-06} Malakis A, Kalozoumis P and Tyraskis N, 2006 \emph{Eur. Phys. J. B} {\bf 50} 63

\bibitem{wu-82} Wu F Y, 1982 \emph{Rev. Mod. Phys.} {\bf 54} 235

\bibitem{monroe-07} Monroe J L and Kim S I, 2007 \emph{Phys. Rev. E} {\bf 76} 021123

\bibitem{order} For the definition of the order-parameter we follow
reference~\cite{malakis-06}, using the four sublattice
magnetizations: $M=\sum_{i=1}^{4}|M_{i}|/4$

\bibitem{ferrenberg-91} Ferrenberg A M and Landau D P, 1991 \emph{Phys. Rev. B} {\bf 44} 5081

\bibitem{kim-96} Kim J -K, 1996 \emph{Phys. Rev. B} {\bf 53} 3388

\bibitem{holm-94} Holm C and Janke W, 1994 \emph{J. Phys. A} {\bf 27} 2553

\bibitem{picco-96} Picco M, 1996 \emph{Phys. Rev. B} {\bf 96} 14930

\bibitem{fisch-78} Fisch R, 1978 \emph{J. Stat. Phys.} {\bf 18} 111

\bibitem{kinzel-81} Kinzel W and Domany E, 1981 \emph{Phys. Rev. B} {\bf 23} 3421

\end{thebibliography}
\end{document}